\documentclass[journal,12pt, draftcls,onecolumn ]{IEEEtran}
\usepackage[top=1in, bottom=1in, left=1in, right=1in]{geometry}
\usepackage{cite}
\usepackage{footnote}
\usepackage{amsmath,amsfonts,amssymb,graphicx}
\usepackage{graphicx}
\usepackage{ifpdf}
\usepackage{amsmath}
\ifCLASSINFOpdf
\else
\fi

\begin{document}
\title{Estimation in Phase-Shift and Forward \\
Wireless Sensor Networks}

\author{\IEEEauthorblockN{Feng Jiang, Jie Chen, and A. Lee Swindlehurst\thanks{This work is supported by the Air Force Office of Scientific
Research grant FA9550-10-1-0310, and by the National Science Foundation
under grant CCF-0916073.}}\\
\IEEEauthorblockA{Department of Electrical Engineering and Computer Science\\University of California at Irvine,\\ Irvine, CA 92697, USA\\Email:\{feng.jiang, jie.chen, swindle\}@uci.edu }}


%


\maketitle

\begin{abstract}
We consider a network of single-antenna sensors that observe an
unknown deterministic parameter.  Each sensor applies a phase shift to
the observation and the sensors simultaneously transmit the result to
a multi-antenna fusion center (FC).  Based on its knowledge of the
wireless channel to the sensors, the FC calculates values for the
phase factors that minimize the variance of the parameter estimate,
and feeds this information back to the sensors.  The use of a
phase-shift-only transmission scheme provides a simplified analog
implementation at the sensor, and also leads to a simpler algorithm
design and performance analysis.  We propose two algorithms for this
problem, a numerical solution based on a relaxed semidefinite
programming problem, and a closed-form solution based on the analytic
constant modulus algorithm.  Both approaches are shown to provide
performance close to the theoretical bound.  We derive asymptotic
performance analyses for cases involving large numbers of sensors or
large numbers of FC antennas, and we also study the impact of phase
errors at the sensor transmitters.  Finally, we consider the sensor
selection problem, in which only a subset of the sensors is chosen to
send their observations to the FC.

\begin{keywords}
Wireless sensor networks, analog sensor networks, distributed beamforming,
phase-only beamforming, sensor management
\end{keywords}

\end{abstract}


%

\newpage
\section{Introduction}
\subsection{Background}

Wireless sensor networks (WSNs) have been widely studied for detection
and estimation problems. Recently, considerable research has focused
on the fusion of \emph{analog} rather than encoded digital data in a
distributed sensor network to improve estimation performance.  The
advantages of analog WSNs have been established in
\cite{Gastpar:2008,Gastpar:2005,Gastpar:2003}, where it was shown that
when using distortion between the source and recovered signal as the
performance metric, digital transmission (separate source and channel
coding) achieves an exponentially worse performance than analog
signaling.  A number of studies have focused on algorithm development
and analysis for analog WSNs with a single-antenna fusion center (FC).
In \cite{Cui:2007}, the sensors amplify and forward their observations
of a scalar source to the FC via fading channels, and algorithms are
developed to either minimize estimation error subject to transmit
power constraints or minimize power subject to estimation error
constraints.  The scalar source model for this problem was generalized
to correlated vector sources in \cite{Bahceci:2008}.  An opportunistic
power allocation approach was proposed in \cite{Matamoros:2011}, and
the scaling law with respect to the number of sensors was shown to be
the same as the optimal power allocation proposed in
\cite{Cui:2007}. In \cite{Banavar:2010}, the asymptotic variance of
the best linear unbiased estimator of an analog WSN is derived,
together with an analysis of the effect of different assumptions
regarding channel knowledge at the sensors.  Scaling laws with respect
to the number of sensors have been studied in \cite{Leong:2010} for a
diversity-based method (where only the sensor with the best channel
transmits), as well as for the coherent multiple access channel (MAC)
and orthogonal channel cases, assuming a Gaussian source.  In
\cite{Wang:2011}, a power optimization problem was formulated to
minimize the outage probability of the MSE for the coherent MAC
channel. More complicated settings involving analog WSNs with
nonlinear measurement models \cite{Fang:2009} or relays
\cite{Thatte:2008, Zarifi:2011} have also been studied.

The results described above all assume that the FC is equipped with
only one antenna. Just as multi-antenna receivers can provide
significant capacity or diversity gains in communication systems, the
estimation performance of a WSN should also benefit from the use of a
multi-antenna FC, though prior work on this scenario is limited.  A
general scenario is investigated in \cite{Xiao:2008}, involving vector
observations of a vector-valued random process at the sensors, and
linearly precoded vector transmissions from the sensors to a
multi-antenna FC.  Optimal solutions for the precoders that minimize
the mean-squared error (MSE) at the FC are derived for a coherent MAC
under power and bandwidth constraints.  In \cite{Smith:2009},
single-antenna sensors amplify and forward their observations to a
multi-antenna FC, but it is shown that for Rayleigh fading channels, the
improvement in estimate variance is upper bounded by only a factor of
two compared to the case of a single-antenna FC.  The performance of
two heuristic algorithms for choosing the gain and phase of the sensor
transmissions is also studied.  Subsequent results by the same
authors in \cite{Banavar:2012, Banavar:20102}, have demonstrated that
when the channel undergoes (zero-mean) Rayleigh fading, there is a
limit to the improvement in detection performance for a multi-antenna
FC as well, but when the channel is Rician, performance improves
monotonically with respect to number of antennas.

The term ``amplify and forward'' is often used to describe analog
sensor networks like those discussed above, since each sensor applies
a complex gain to the observation before sending it to the FC.  For a
coherent MAC, one can think of this as a type of distributed transmit
beamforming, although it is distinguished from distributed beamforming
applications such as those in communications since in a WSN the
observed noise is transmitted together with the signal of interest.
Some prior research in radar and communications has focused on
scenarios where the beamformer weights implement only a phase
shift rather than both a gain and a phase.  The advantage of using
phase shifting only is that it simplifies the implementation and is
easily performed with analog hardware.  Phase-shift-only beamformers
have most often been applied to receivers that null spatial
interference \cite{Smith:1999, Kajenski:2012}, but it has also been
considered on the transmit side for MISO wireless communications
systems \cite{Xia:2009}, which is similar to the problem considered
here.  For the distributed WSN estimation problem, phase-only sensor
transmissions have been proposed in \cite{Tepedelenlioglu:2010}, where
the phase is a scaled version of the observation itself. Phase-only
transmissions were also considered in the context of distributed
detection in \cite{Banavar:2012}, leading to a problem similar to
one of those we consider here.

In addition to the work outlined above, other WSN research has focused
on sensor selection problems, particularly in situations where the
sensors have limited battery power.  In these problems, only a subset
of the sensors are chosen to transmit their observations, while the
others remain idle to conserve power.  The sensor selection problem
has been tackled from various perspectives, with the goal of
optimizing the estimation accuracy
\cite{Thatte:2008,Joshi:2009,Gupta:2006} or some heuristic system
utility \cite{Fang:2006,Krishnamurthy:2008}.  In \cite{Joshi:2009},
the authors investigated maximum likelihood (ML) estimation of a
vector parameter by selecting a fixed-size subset of the sensors. An
approximate solution was found by relaxing the original Boolean
optimization to a convex optimization problem.  A dynamic model is
used to describe the parameter of interest in \cite{Gupta:2006}, and
sensors use the Kalman filter to estimate the parameter. At each time
step, a single sensor is selected and the measurement at the selected
sensor is shared with all other sensors. A numerical sensor selection
algorithm was proposed to minimize an upper bound on the expected
estimation error covariance. Instead of the estimation accuracy, a
utility function that takes into account the measurement quality or
energy cost can also be used as the metric for sensor selection. In
\cite{Krishnamurthy:2008}, each sensor independently optimizes its own
operation status based on a utility function which depends on the
sensor's own measurement and the predicted operation status of other
sensors. A threshold is then found to enable the sensor to switch its
status for either energy efficiency or energy consumption, and a power
allocation algorithm was proposed to minimize the MSE at FC.

\subsection{Approach and Contributions}
In this paper we consider a distributed WSN with single-antenna
sensors that observe an unknown deterministic parameter corrupted by
noise. The low-complexity sensors apply a phase shift (rather than
both a gain and phase) to their observation and then simultaneously
transmit the result to a multi-antenna FC over a coherent MAC.  One
advantage of a phase-shift-only transmission is that it leads to a
simpler analog implementation at the sensor.  The FC determines the
optimal value of the phase for each sensor in order to minimize the ML
estimation error, and then feeds this information back to the sensors
so that they can apply the appropriate phase shift. The estimation
performance of the phase-optimized sensor network is shown to be
considerably improved compared with the non-optimized case, and close
to that achieved by sensors that can adjust both the transmit gain and
phase.  We analyze the asymptotic behavior of the algorithm for a
large number of sensors and a large number of antennas at the FC.  In
addition, we analyze the impact of phase errors at the sensors due,
for example, to errors in the feedback channel, a time-varying main
channel or phase-shifter drift.  We also consider a sensor selection
problem similar to that in \cite{Joshi:2009}, and analyze its
asymptotic behavior as well.  Some additional details regarding the
contributions of the paper are listed below.
\begin{enumerate}
\item We present two algorithms for determining the phase factors used
at each sensor.  In the first, we use the semi-definite relaxation
presented in \cite{Luo:2006,Banavar:2012} to convert the original problem to a
semidefinite programming (SDP) problem that can be efficiently solved
by interior-point methods.  For the second algorithm, we apply the
analytic constant modulus algorithm (ACMA) \cite{Alle:1996}, which provides a
considerably simpler closed-form solution.  Despite the reduction
in complexity, the performance of ACMA is shown via simulation to be
only slightly worse than the SDP solution, and close to the theoretical
lower bound on the estimate variance.  This is especially encouraging
for networks with a large number of sensors $N$, since the SDP complexity
is on the order of $N^{3.5}$, while that for ACMA is only on the order of
$N^2$.
\item We separately derive performance scaling laws with respect to
the number of antennas and the number of sensors assuming non-fading
channels that take path loss into account.  For both cases, we derive conditions that
determine whether or not the presence of multiple antennas at the FC
provides a significant benefit to the estimation performance. Prior work in \cite{Smith:2009,Banavar:2012,Banavar:20102}
has focused on either AWGN channels with identical channel gains, or
on fading channels where the channel gains are identically distributed,
corresponding to the case where the distances from the sensors to the
FC are roughly the same.  References \cite{Smith:2009,Banavar:2012,Banavar:20102}
also assume a special case where the noise at each of the sensors
has the same variance, although \cite{Banavar:20102} examines how
certain upper bounds on performance change when the sensor noise is
arbitrarily correlated.
\item Using our model for the non-fading case, we are able
to elucidate detailed conditions under which the asymptotic estimation performance
will improve with the addition of more antennas $M$ at the FC.  While
\cite{Smith:2009,Banavar:2012} showed that performance always improves
with increasing $M$ for AWGN channels with identical gains and identically
distributed sensor noise, we derive more detailed conditions that take
into account the possibility of non-uniform distances between the sensors
and FC and non-uniform noise at the sensors.
\item We conduct an analysis of the impact of phase errors at the
sensors assuming relatively small phase errors with variance
$\sigma_p^2\ll1$ (square-radians).  In particular, we show that the
degradation to the estimate variance is bounded above by a factor of
$1+\sigma_p^2$.  We note that the effect of errors in the transmit
phase at the sensors has previously been considered for the case of
$M=1$ in \cite{Banavar:2010}, although using a different phase
error model.
\item We consider the sensor selection problem separately for low
and high sensor measurement noise.  For the low measurement noise
scenario, we relax the sensor selection problem to a standard linear
programming (LP) problem, and we also propose a reduced complexity version
of the algorithm.  For the high measurement noise scenario, we show
that the estimation error is lower bounded by the inverse of the
measurement noise power, which motivates the use of a simple selection
method based on choosing the sensors with the lowest measurement
noise.
\end{enumerate}
A subset of the above results was presented in an earlier conference
paper \cite{Jiang:2011}.

\subsection{Organization}
The paper is organized as follows. Section~\ref{sec:two} describes the
assumed system model. Section~\ref{sec:three} formulates the phase
optimization problem and proposes a numerical solution based on SDP as
well as a closed-form solution based on the algebraic constant modulus
algorithm.  In Section~\ref{sec:four}, the asymptotic performance of
the algorithm is analyzed for a large number of sensors and antennas.
The effect of phase errors is analyzed in Section~\ref{sec:five} and the
sensor selection problem is investigated in Section~\ref{sec:six}. 
Simulation results are then presented in Section~\ref{sec:seven} and
our conclusions can be found in Section~\ref{sec:eight}.

\section{System Model}\label{sec:two}

We assume that $N$ single-antenna sensors in a distributed sensor
network independently observe an unknown but deterministic complex-valued parameter
$\theta$ according to the following model for sensor $i$:
\begin{equation}
\label{eq:sensmod}
y_i = \theta + v_i \;\nonumber ,
\end{equation}
where $v_i$ is complex-valued Gaussian observation noise with variance $\sigma_{v,i}^2$.
The noise is assumed to be independent from sensor to sensor.
Each sensor phase shifts its observation and transmits the signal
$a_i y_i$ to the FC, where $|a_i|=1$.  Assuming a coherent MAC and an
FC with $M$ antennas, the vector signal received at the FC can be
expressed as
\begin{equation}
\label{eq:yvec}
\mathbf{y}=\mathbf{H}\mathbf{a}\theta+\mathbf{H}\mathbf{D}\mathbf{v}+\mathbf{n}\;,
\end{equation}
where $\mathbf{H}=[\mathbf{h}_{1},\dots,\mathbf{h}_{N}]$ and
$\mathbf{h}_{i}\in\mathbb{C}^{M\times 1}$ is the channel vector
between the $i$th sensor and the FC,
$\mathbf{a}=[a_{1},\dots,a_{N}]^{T}$ contains the adjustable phase
parameters, $\mathbf{D}=\mathrm{diag}\{a_{1},\dots,a_{N}\}$,
$\mathbf{v}$ is the sensor measurement noise vector with covariance
$\mathbf{V}=\mathbb{E}\{\mathbf{v}\mathbf{v}^{H}\}=\mathrm{diag}\left\{\sigma_{v,1}^2,\cdots,\sigma_{v,N}^2\right\}$,
and $\mathbf{n}$ is complex Gaussian noise at the FC with covariance
$\mathbb{E}\{\mathbf{n}\mathbf{n}^{H}\}=\sigma_{n}^2\mathbf{I}_{M}$,
where $\mathbf{I}_{M}$ is an $M\times M$ identity matrix.  Note
that since the sensors can only phase shift their observation prior
to transmission, we ignore the issue of power control and assume
that the sensors have sufficient power to forward their observation
to the FC.

The combined noise term $\mathbf{HDv}+\mathbf{n}$ in~(\ref{eq:yvec})
is Gaussian with covariance $\mathbf{HVH}^H+\sigma_n^2\mathbf{I}$,
since $\mathbf{DVD}^H=\mathbf{V}$ due to the phase-only assumption.
Assuming the FC is aware of the channel matrix
$\mathbf{H}$, the noise covariance $\mathbf{V}$ and $\sigma_n^2$,
it can calculate the ML estimate of $\theta$ using \cite{Kay:1993}
\begin{equation}
\hat{\theta}_{ML}=\frac{\mathbf{a}^{H}\mathbf{H}^{H}(\mathbf{H}\mathbf{V}\mathbf{H}^{H}+\sigma_{n}^2\mathbf{I}_{M})^{-1}\mathbf{y}}{\mathbf{a}^{H}\mathbf{H}^{H}(\mathbf{H}\mathbf{V}\mathbf{H}^{H}+\sigma_{n}^2\mathbf{I}_{M})^{-1}\mathbf{H}\mathbf{a}}\;.\nonumber
\end{equation}
The estimator $\hat{\theta}_{ML}$ is unbiased with variance
\begin{equation}\label{eq:ml}
\mathrm{Var}(\hat{\theta}_{ML})=\left(\mathbf{a}^{H}\mathbf{H}^{H}(\mathbf{H}\mathbf{V}\mathbf{H}^{H}+\sigma_{n}^2\mathbf{I}_{M})^{-1}\mathbf{H}\mathbf{a}\right)^{-1}\;.
\end{equation}
Furthermore, since $\|\mathbf{a}\|=N$ when only phase shifts are used
at the sensors, it is easy to see that the variance is lower bounded by
\begin{equation}\label{eq:lb}
\mathrm{Var}(\hat{\theta}_{ML})\!\ge\!\frac{1}{N\lambda_{\max}\left(\mathbf{H}^{H}(\mathbf{H}\mathbf{V}\mathbf{H}^{H}+\sigma_{n}^2\mathbf{I}_{M})^{-1}\mathbf{H}\right)}\;,
\end{equation}
where $\lambda_{\max}(\cdot)$ denotes the largest eigenvalue of its
matrix argument.  Note that the bound in~(\ref{eq:lb}) is in general
unachievable, since with probability one the given matrix will not
have an eigenvector with unit modulus elements.

\section{Optimizing the Sensor Phase}\label{sec:three}

In this section we consider the problem of choosing $\mathbf{a}$ to
minimize $\mathrm{Var}(\hat{\theta}_{ML})$ in~(\ref{eq:ml}).  The unit
modulus constraint prevents a trivial solution, but as we note below,
a direct solution is not possible even without this constraint since
the noise covariance would then depend on $\mathbf{a}$.  The general
optimization problem is formulated as
\begin{eqnarray}\label{eq:opt}
\min_{\mathbf{a}} &&\mathrm{Var}(\hat{\theta}_{ML})\\
s. t. &&|a_i|=1,\; i=1, \dots, N\;.\nonumber
\end{eqnarray}
Defining
$\mathbf{B}=\mathbf{H}^{H}(\mathbf{H}\mathbf{V}\mathbf{H}^{H}+\sigma_{n}^2\mathbf{I}_{M})^{-1}\mathbf{H}$,
the problem can be rewritten as
\begin{eqnarray}\label{eq:quardra}
\max_{\mathbf{a}} &&\mathbf{a}^{H}\mathbf{B}\mathbf{a}\\
s. t. &&|a_i|=1,\; i=1, \dots, N\;.\nonumber
\end{eqnarray}
Note that this optimization can only determine $\mathbf{a}$ to
within an arbitrary phase shift $e^{j\phi}$, but this scaling has no
impact on the estimate of $\theta$.  In other words, the vector $\mathbf{a}$
and the vector $\mathbf{a}e^{j\phi}$ for arbitrary $\phi$ will both yield
the same estimate $\hat{\theta}_{ML}$.  Since the FC is aware of the vector
$\mathbf{a}$ determined by the optimization in~(\ref{eq:quardra}), any
arbitrary phase factor present in the $\mathbf{Ha}\theta$ term of the
model in~(\ref{eq:yvec}) will be canceled when the ML estimate of $\theta$
is computed.  This is also clear from the variance expression in~(\ref{eq:ml}),
which is insensitive to any phase shift to $\mathbf{a}$.

If there are only two sensors in the network, a simple closed-form
solution to~(\ref{eq:quardra}) can be obtained.  Defining
$\mathbf{B}=\left[\begin{array}{cc} a&be^{j\beta}\\
be^{-j\beta}&c\end{array}\right]$ with $a, b, c>0$ and $\mathbf{a}=[e^{j\beta_1},e^{j\beta_2}]$, then $\mathbf{a}^H\mathbf{B}\mathbf{a}$ is calculated as
\begin{eqnarray}\label{eq:closetwo}
\mathbf{a}^H\mathbf{B}\mathbf{a}&=&a+c+2b\cos(\beta_1-\beta_2-\beta)\nonumber\\
&\le&a+c+2b\;,
\end{eqnarray}
and the equality in (\ref{eq:closetwo}) can be achieved for any
$\beta_1,\beta_2$ that satisfy $\beta_1-\beta_2=\beta$.
For the general situation where $N>2$, a solution to~(\ref{eq:quardra})
appears to be intractable.  Instead, in the discussion that follows
we present two suboptimal approaches in order to obtain an approximate
solution.  The first approach is based on an SDP problem obtained by
relaxing a rank constraint in a reformulated version of~(\ref{eq:quardra}),
similar to the approach proposed in \cite{Luo:2006,Banavar:2012}.
The second converts the problem to one that can be solved via the
ACMA of \cite{Alle:1996}.
It is worth emphasizing here that if the transmission gain of the
sensors was also adjustable, then the corresponding problem would be
\begin{eqnarray}
\max_{\mathbf{a}} && \mathbf{a}^{H}\mathbf{H}^{H}(\mathbf{H}\mathbf{DVD}^H\mathbf{H}^{H}
+\sigma_{n}^2\mathbf{I}_{M})^{-1}\mathbf{H}\mathbf{a} \label{eq:amplitudeopt} \\
s. t. && \mathbf{a}^H\mathbf{a} \le N \; , \nonumber
\end{eqnarray}
which also has no closed-form solution due to the dependence on
$\mathbf{a}$ (through the matrix $\mathbf{D}$) inside the matrix
inverse.  While in general both our SDP solution and~(\ref{eq:amplitudeopt})
require numerical
optimizations, we will see in Sections~IV-VI that the theoretical
analysis of performance and the solution to the sensor selection
problem is considerably simpler with the phase-only constraint.
The simulations of Section~VII will also demonstrate that there is
often little performance loss incurred by using phase-shift-only
transmissions.

\subsection{SDP Formulation}
To begin, we rewrite~(\ref{eq:quardra}) as follows:
\begin{eqnarray}\label{eq:quardra2}
\max_{\mathbf{a}} &&\mathrm{tr}\left(\mathbf{B}\mathbf{a}\mathbf{a}^{H}\right)\\
s. t. &&|a_i|=1,\; i=1, \dots, N\;.\nonumber
\end{eqnarray}
Making the association $\mathbf{A}=\mathbf{a}\mathbf{a}^H$,
problem~(\ref{eq:quardra2}) is equivalent to:
\begin{eqnarray}\label{eq:rankone}
\max_{\mathbf{A}} && \mathrm{tr}(\mathbf{B}\mathbf{A})\\
s. t. &&\mathbf{A}_{i,i}=1,\; i=1, \dots, N\nonumber\\
      && \mathrm{rank}(\mathbf{A})=1\nonumber\\
      &&\mathbf{A}\succeq 0\nonumber\;,
\end{eqnarray}
where $\mathbf{A}_{i,i}$ denotes the $i$th diagonal element of
$\mathbf{A}$.  Following the approach of \cite{Luo:2006,Banavar:2012},
we then relax the rank-one constraint, so that
the problem becomes a standard SDP:
\begin{eqnarray}\label{eq:sdp}
\max_{\mathbf{A}} && \mathrm{tr}(\mathbf{B}\mathbf{A})\\
s. t. &&\mathbf{A}_{i,i}=1,\; i=1, \dots, N\nonumber\\
      &&\mathbf{A}\succeq 0\nonumber\;.
\end{eqnarray}
Defining $\mathbf{B}_r=\mbox{\rm real}\{\mathbf{B}\}$,
$\mathbf{B}_i=\mbox{\rm imag}\{\mathbf{B}\}$, and similarly for
$\mathbf{A}_r$ and $\mathbf{A}_i$, we can convert~(\ref{eq:sdp}) to
the equivalent real form
\begin{eqnarray}\label{eq:realsdp}
\max_{\{\mathbf{A}_r,\mathbf{A}_i\}} && \mathrm{tr}(\mathbf{B}_r\mathbf{A}_r-\mathbf{B}_i\mathbf{A}_i)\\
s. t. &&\mathbf{A}_{r~i,i}=1,\; i=1, \dots, N\nonumber\\
      &&\left[\begin{array}{cc} \mathbf{A}_r&-\mathbf{A}_i\\
                \mathbf{A}_i&\mathbf{A}_r\end{array}\right]\succeq 0\nonumber\;.
\end{eqnarray}
Problem (\ref{eq:realsdp}) can be efficiently solved by a standard
interior-point method \cite{Boyd:2004}.

In general, the solution to~(\ref{eq:realsdp}) will not be rank one,
so an additional step is necessary to estimate $\mathbf{a}$.
Let $\mathbf{A}_r^*$, $\mathbf{A}_i^*$ denote the solution to problem (\ref{eq:realsdp}), then the solution to problem (\ref{eq:sdp}) is given by $\mathbf{A}^{*}=\mathbf{A}_r^*+j\mathbf{A}_i^*$. If $\mathrm{rank}(\mathbf{A}^{*})>1$, we can use a
method similar to Algorithm 2 in \cite{Zhang:2011} to extract a
rank-one solution, as follows:
\begin{enumerate}
\item Decompose\footnote{Since $\mathbf{A}^{*}$ is the solution to
problem (\ref{eq:sdp}), $\mathbf{A}^{*}$ is positive semidefinite.}
$\mathbf{A}^{*}=\mathbf{C}^{H}\mathbf{C}$, define
$\tilde{\mathbf{B}}=\mathbf{C}\mathbf{B}\mathbf{C}^{H}$, and find a
unitary matrix $\mathbf{U}$ that can diagonalize $\tilde{\mathbf{B}}$.
\item Let $\mathbf{r}\in\mathbb{C}^{N\times 1}$ be a random vector
whose $i$th element is set to $e^{j\omega_i}$, where
$\omega_i$ is uniformly distributed over $[0, 2\pi)$.
\item Set $\tilde{\mathbf{a}}=\mathbf{C}^{H}\mathbf{U}\mathbf{r}$, and
the solution is given by $\mathbf{a}^*=[a_1^* \; \cdots \; a_N^*]^T$,
where $a_i^* = e^{j\angle{\tilde{a}_i}}$ and $\angle{z}$ represents
the phase of a complex number $z$.
\end{enumerate}
A detailed discussion of the reasoning behind the above rank-one modification
can be found in \cite{Zhang:2011}.

\subsection{ACMA Formulation}

For this discussion, we will assume that $N > M$, which represents the
most common scenario.  Thus, the $N\times N$ matrix $\mathbf{B}$ in
the quadratic form $\mathbf{a}^H\mathbf{Ba}$ that we are trying to
maximize is low rank; in particular, $\mbox{\rm rank}(\mathbf{B}) \le
M < N$.  Clearly, any component of $\mathbf{a}$ orthogonal to the
columns or rows of $\mathbf{B}$ will not contribute to our goal of
minimizing the estimate variance.  In particular, if we define the
singular value decomposition (SVD)
$\mathbf{B}=\mathbf{U}\boldsymbol{\Sigma}\mathbf{U}^H$, we ideally
seek a vector $\mathbf{a}$ such that
\begin{eqnarray}
\label{eq:acma1}
\mathbf{a}&=&\sum_{k=1}^m w_k \mathbf{u}_k = \mathbf{U}_m \mathbf{w}\\
|a_i|&=&1 \;,\nonumber
\end{eqnarray}
where $\mathbf{U}_m=[\mathbf{u}_1 \; \cdots \; \mathbf{u}_m]$ contains
the first $m \le \mbox{\rm rank}(\mathbf{B}) \le M$ singular vectors
of $\mathbf{B}$ and $\mathbf{w}=[w_1 \; \cdots \; w_m]^T$.  The
problem of finding the coefficient vector $\mathbf{w}$ of a linear
combination of the columns of a given matrix $\mathbf{U}_m$ that
yields a vector with unit modulus elements is precisely the problem
solved by the ACMA \cite{Alle:1996}.

Our problem is slightly different from the one considered in
\cite{Alle:1996}, since there will in general be no solution
to~(\ref{eq:acma1}) even in the absence of noise.  However, in our
simulation results we will see that the ACMA solution provides
performance close to that obtained by the SDP formulation above.  Note
also that there is a trade-off in the choice of $m$, the number of
vectors in $\mbox{\rm span}(\mathbf{B})$ to include in the linear
combination of~(\ref{eq:acma1}).  A small value of $m$ allows us to focus on forming
$\mathbf{a}$ from vectors that will tend to increase the value of
$\mathbf{a}^H\mathbf{Ba}$, while a larger value for $m$ provides more
degrees of freedom in finding a vector whose elements satisfy
$|a_i|=1$.  Another drawback to choosing a larger value for $m$ is
that the ACMA solution can only be found if $N > m^2$.  As long as $M$
is not too large, one could in principle try all values of
$m=1,\cdots,M$ that satisfy $N > m^2$ and choose the one that yields
the smallest estimate variance.  We will see later in the simulations
that a small value for $m$ already provides good performance, so the
choice of $m$ is not a significant issue.

The general ACMA approach can be formulated to find multiple solutions
to~(\ref{eq:acma1}), but in our case we only need a single solution, and
thus a simplified version of ACMA can be used, as outlined here for a given $m$.
The ACMA is obtained by defining the rows of $\mathbf{U}_m$ as
$\mathbf{U}_m^H=[\tilde{\mathbf{u}}_1 \; \cdots \;\tilde{\mathbf{u}}_N]$,
and then rewriting the constraint  $|a_i|=|\tilde{\mathbf{u}}_i^H \mathbf{w}| = 1$ as
\[ \left(\bar{\tilde{\mathbf{u}}}_i \otimes \tilde{\mathbf{u}}_i\right)^H
\left(\bar{\mathbf{w}} \otimes \mathbf{w}\right) = 1 \; ,
\]
where $\bar{(\cdot)}$ denotes the complex conjugate and $\otimes$ the
Kronecker product.  Stacking all
$N$ such constraints into a single equation results in
\begin{equation}\label{eq:Pw}
\mathbf{P} \left(\bar{\mathbf{w}} \otimes \mathbf{w}\right) = 0 \; ,
\end{equation}
where
\begin{equation}
\mathbf{P} = \left[ \begin{array}{cc} \left(\bar{\tilde{\mathbf{u}}}_1
\otimes \tilde{\mathbf{u}}_1\right)^H & -1 \\ \vdots & \vdots \\
\left(\bar{\tilde{\mathbf{u}}}_N
\otimes \tilde{\mathbf{u}}_N\right)^H & -1 \end{array} \right] \; .
\end{equation}
If an exact solution to~(\ref{eq:Pw}) existed, then a vector in the
null space of $\mathbf{P}$ would have the form $\left[
\left(\bar{\mathbf{w}} \otimes \mathbf{w}\right)^T \; \; 1 \right]^T$, and
$\mathbf{w}$ could be found by stripping away the $1$ and then
unstacking the resulting vector into a rank-one matrix (see \cite{Alle:1996}
for more details).  In our problem, an exact solution to~(\ref{eq:Pw})
does not exist, so we use the following approach to obtain an approximation:
\begin{enumerate}
\item Let $\mathbf{q}$ represent the right singular vector of $\mathbf{P}$
associated with the smallest singular value, and define the vector
$\tilde{\mathbf{q}}$ to contain the first $m^2$ elements of $\mathbf{q}$.
\item Set $\mathbf{w}$ equal to the singular vector of $\tilde{\mathbf{Q}}
+ \tilde{\mathbf{Q}}^H$ with largest singular value, where the $m \times m$
matrix
\begin{equation}
\tilde{\mathbf{Q}} = \mbox{\rm vec}^{-1} (\tilde{\mathbf{q}})
\end{equation}
is formed by dividing $\tilde{\mathbf{q}}$ into sub-vectors of length
$m$ and stacking them together in a matrix.
\item Set $\hat{\mathbf{a}} = \mathbf{U}_m \mathbf{w}$.  The
vector $\mathbf{a}$ is then found by setting the magnitude of all
the elements of $\hat{\mathbf{a}}$ equal to unity.  In particular,
the $i$-th element of $\mathbf{a}$ is given by
\begin{equation*}
a^*_i = e^{j\angle{\hat{a}_i}} \; .
\end{equation*}
\end{enumerate}

\subsection{Comparison of Computational Complexity}

As discussed in \cite{Luo:2006}, the computational load of the SDP problem
in~(\ref{eq:sdp}) is of the order $O(N^{3.5})$.  The additional steps
required to take the SDP result and find a rank-one solution require an
$O(N^3)$ eigenvalue decomposition, so the overall complexity is dominated
by the SDP.  For ACMA, the dominant computational step occurs in finding the
$m$ principal eigenvectors of the Hermitian matrix $\mathbf{B}$, which 
requires only an order $O(mN^2)$ computation \cite{Golub:1989}.  Finding the 
least dominant singular vector of $\mathbf{P}$ is an $O(N^2)+O(m^4)$ operation, 
and the remaining steps have relatively trivial complexity.  Since $m \ll N$
in typical scenarios, we see that ACMA enjoys a significantly lower computational
load compared to the SDP approach.  Despite this, we will see that ACMA has
performance that is only slightly inferior to using the SDP solution.

\section{Asymptotic Performance Analysis}\label{sec:four}

In this section, we analyze the asymptotic performance achievable
using only phase-shifts for the sensor transmissions.  We will
separately study cases where the number of sensors is large ($N
\rightarrow\infty$) or the number of FC antennas is large
($M\rightarrow\infty$).  Our analysis will be based on an a non-fading
channel model that takes path loss into account, similar to models
used in \cite{Gerhard:2003, Jafar:2011}.  In particular, for the
channel between the FC and sensor $i$, we assume
\begin{equation}
\mathbf{h}_{i}=\frac{1}{d_i^\alpha}\tilde{\mathbf{h}}_{i}\;,\nonumber
\end{equation}
where $d_i$ denotes the distance between the $i$th sensor and the FC,
$\alpha$ is the path loss exponent and $\tilde{\mathbf{h}}_i$ is
given by
\begin{equation}
\tilde{\mathbf{h}}_{i}=[e^{j\gamma_{i,1}} \; \; e^{j\gamma_{i,2}} \; \cdots \; e^{j\gamma_{i,M}}]^{T}\nonumber\;,
\end{equation}
where $\gamma_{i,j}$ is uniformly distributed over $\left[0,2\pi\right)$.

\subsection{Estimation Performance for Large $N$}

From (\ref{eq:lb}) we know that the lower bound on
$\mathrm{Var}(\hat{\theta}_{ML})$ depends on the largest eigenvalue of
$\mathbf{H}^{H}(\mathbf{H}\mathbf{V}\mathbf{H}^{H}+\sigma_{n}^2\mathbf{I}_{M})^{-1}\mathbf{H}$. 
We begin by deriving a lower bound for this eigenvalue.
The $(m,n)$th element of $\mathbf{H}\mathbf{V}\mathbf{H}^H$ can be expressed as
\begin{eqnarray}
\left(\mathbf{H}\mathbf{V}\mathbf{H}^{H}\right)_{m,n}=\sum_{i=1}^{N}
\frac{e^{j(\gamma_{i,m}-\gamma_{i,n})}\sigma_{v,i}^2}{d_i^{2\alpha}}\;.\nonumber
\end{eqnarray}
According to the strong law of large numbers, as $N\to \infty$ we have
\begin{eqnarray}\label{eq:mean}
\lim_{N\to\infty}\frac{1}{N}\sum_{i=1}^{N}\frac{e^{j(\gamma_{i,m}\!-\gamma_{i,n})}\sigma_{v,i}^2}{d_i^{2\alpha}}&\overset{(a)}{=}&\mathbb{E}\left\{\frac{\sigma_{v,i}^2}{d_i^{2\alpha}}\right\}\mathbb{E}\left\{e^{j(\gamma_{i,m}-\gamma_{i,n})}\right\}\nonumber 
\\ &\overset{(b)}{=}&\left\{\begin{array}{lr}
\mathbb{E}\left\{\frac{\sigma_{v,i}^2}{d_i^{2\alpha}}\right\} & m=n\\
0 & m\ne n \; ,
\end{array}\right.
\end{eqnarray}
where ($a$) follows from the assumption that $\gamma_{i,m}$, $d_i$ and
$\sigma_{v,i}^2$ are independent  and ($b$) is due to
the fact that $\gamma_{i,m}$ and $\gamma_{i,n}$ are independent and
uniformly distributed over $[0,2\pi)$. Thus, for sufficiently
large $N$ we have
\begin{equation}\label{eq:approx}
\lim_{N\to\infty}\mathbf{H}\mathbf{V}\mathbf{H}^{H}=N\mathbb{E}\left\{\frac{\sigma_{v,i}^2}{d_i^{2\alpha}}\right\}\mathbf{I}_{M} \; .
\end{equation}

Based on (\ref{eq:approx}), we have
\begin{eqnarray}\label{eq:lambdamax}
\lim_{N\to\infty}
\lambda_{\max}\left(\mathbf{H}^{H}(\mathbf{H}\mathbf{V}\mathbf{H}^{H}+\sigma_{n}^2\mathbf{I}_{M})^{-1}\mathbf{H}\right)&
= & \frac{1}{N\mathbb{E}\left\{\frac{\sigma_{v,i}^2}{d_i^{2\alpha}}\right\}+\sigma_n^2}
\left[ \lim_{N\to\infty} \lambda_{\max}(\mathbf{H}^H\mathbf{H}) \right]\nonumber\\
&\overset{(c)}{=}&\frac{N\mathbb{E}\left\{\frac{1}{d_i^{2\alpha}}\right\}}{N\mathbb{E}
\left\{\frac{\sigma_{v,i}^2}{d_i^{2\alpha}}\right\}+\sigma_n^2} \; ,
\end{eqnarray}
where ($c$) is due to the fact that
$\lambda_{\max}(\mathbf{H}^{H}\mathbf{H})=\lambda_{\max}(\mathbf{H}\mathbf{H}^{H})$. Substituting
(\ref{eq:lambdamax}) into (\ref{eq:lb}), we have the following asymptotic
lower bound on the estimate variance:
\begin{equation}\label{eq:lb2}
\mathrm{Var}(\hat{\theta}_{ML}) \ge
\frac{N\mathbb{E}\left\{\frac{\sigma_{v,i}^2}{d_i^{2\alpha}}\right\}+\sigma_n^2}{N^2\mathbb{E}\left\{\frac{1}{d_i^{2\alpha}}\right\}}
\; .
\end{equation}
For large enough $N$, the lower bound can be approximated using sample
averages:
\begin{equation}\label{eq:lb2app}
\mathrm{Var}(\hat{\theta}_{ML})\ge\frac{\sum_{i=1}^{N}
\frac{\sigma_{v,i}^2}{d_i^{2\alpha}}+\sigma_{n}^2}{N\sum_{i=1}^{N}\frac{1}{d_i^{2\alpha}}} \; .
\end{equation}

Next, we derive an upper bound on the estimate variance and compare it
with the lower bound obtained above.  The upper bound is obtained by
calculating the variance obtained when only a single antenna is present
at the FC.  For the given channel model, the optimal choice for the
vector of sensor phases is just the conjugate of the channel phases:
$\mathbf{a}=[e^{-j\gamma_{1,1}} \; \cdots \; e^{-j\gamma_{N,1}}]^T$,
which when applied to~(\ref{eq:ml}) leads to
\begin{equation}\label{eq:allone}
\mathrm{Var}(\hat{\theta}_{ML})\le\frac{\sum_{i=1}^{N}
\frac{\sigma_{v,i}^2}{d_i^{2\alpha}}+\sigma_{n}^2}{\left(\sum_{i=1}^{N}\frac{1}{d_i^\alpha}\right)^2} \; .
\end{equation}
When $N\to\infty$, both the upper and lower bounds converge to $0$,
but the ratio of the lower bound
in~(\ref{eq:lb2app}) to the upper bound in~(\ref{eq:allone})
converges to
\begin{eqnarray}\label{eq:ratio}
\lim_{N\to\infty}\frac{\left(\sum_{i=1}^{N}\frac{1}{d_i^\alpha}\right)^2}{N\sum_{i=1}^{N}\frac{1}{d_i^{2\alpha}}}=\frac{\left(\mathbb{E}\left\{\frac{1}{d_{i}^\alpha}\right\}\right)^2}{\mathbb{E}\left\{\frac{1}{d_{i}^{2\alpha}}\right\}}=1-\frac{\mathrm{Var}\left\{\frac{1}{d_{i}^\alpha}\right\}}{\mathbb{E}\left\{\frac{1}{d_{i}^{2\alpha}}\right\}}\;.
\end{eqnarray}
Interestingly, we see that if
$\mathrm{Var}\left\{\frac{1}{d_{i}^\alpha}\right\}\ll
\mathbb{E}\left\{\frac{1}{d_{i}^{2\alpha}}\right\}$, the gap between
the upper and lower bound is very small, and the availability of
multiple antennas at the FC does not provide much benefit compared
with the single antenna system when $N \rightarrow\infty$.  On the
other hand, if $\mathrm{Var}\left\{\frac{1}{d_{i}^\alpha}\right\}\rightarrow
\mathbb{E}\left\{\frac{1}{d_{i}^{2\alpha}}\right\}$, the potential
exists for multiple antennas to significantly lower the estimate
variance.

\subsection{Estimation Performance for Large $M$}

Using the matrix inversion lemma, we have
\begin{eqnarray}\label{eq:expansion}
\mathbf{H}^{H}(\mathbf{H}\mathbf{V}\mathbf{H}^{H}+\sigma_{n}^2\mathbf{I}_{M})^{-1}\mathbf{H}&
= &
\mathbf{H}^{H}\left(\!\frac{1}{\sigma_n^2}\mathbf{I}_{M}\!-\!\frac{1}{\sigma_n^4}\mathbf{H}\left(\mathbf{V}^{-1}+\frac{1}{\sigma_n^2}\mathbf{H}^{H}\mathbf{H}\right)^{-1}\mathbf{H}^{H}\right)\mathbf{H}\nonumber\\
&=&\frac{1}{\sigma_n^2}\mathbf{H}^{H}\mathbf{H}-\frac{1}{\sigma_n^4}\mathbf{H}^{H}\mathbf{H}\left(\mathbf{V}^{-1}+\frac{1}{\sigma_n^2}\mathbf{H}^{H}\mathbf{H}\right)^{-1}\mathbf{H}^{H}\mathbf{H} \; .
\end{eqnarray}
Furthermore, the $(m,n)$th element of $\mathbf{H}^{H}\mathbf{H}$ is given by
\begin{equation}\label{eq:approxh}
\left(\mathbf{H}^{H}\mathbf{H}\right)_{m,n}=\frac{1}{d_m^\alpha d_n^{\alpha}}\sum_{i=1}^{M}e^{j\left(\gamma_{n,i}-\gamma_{m,i}\right)}\;.
\end{equation}
Similar to (\ref{eq:mean}), as $M\to\infty$ we have
\begin{equation}\label{eq:mean2}
\lim_{M\to\infty}\frac{1}{M}\sum_{i=1}^{M}e^{j\left(\gamma_{n,i}-\gamma_{m,i}\right)}
=\left\{\begin{array}{lr}
1 & m=n\\
0 & m\ne n \; ,
\end{array}\right.
\end{equation}
and thus
\begin{eqnarray}\label{eq:approx2}
\lim_{M\to\infty} \mathbf{H}^{H}\mathbf{H} =
M\mathrm{diag}\left\{\frac{1}{d_1^{2\alpha}} \; \cdots \; \frac{1}{d_N^{2\alpha}}\right\}\;.
\end{eqnarray}
Substituting (\ref{eq:approx2}) into (\ref{eq:expansion}), we have
\begin{eqnarray}\label{eq:eigen2}
\lim_{M\to\infty} \mathbf{H}^{H}(\mathbf{H}\mathbf{V}\mathbf{H}^{H}+\sigma_{n}^2\mathbf{I}_{M})^{-1}
\mathbf{H} = \mathrm{diag}\left\{\frac{M}{d_1^{2\alpha}\sigma_n^2+M\sigma_{v,i}^2} \; , \; \cdots \; , \; \frac{M}{d_N^{2\alpha}\sigma_n^2+M\sigma_{v,N}^2}\right\}\; , \nonumber
\end{eqnarray}
and thus
\begin{equation}\label{eq:varapprox}
\lim_{M\to\infty} \mathrm{Var}(\hat{\theta}_{ML}) =
\frac{1}{M\sum_{i=1}^{N}\frac{1}{d_i^{2\alpha}\sigma_n^2+M\sigma_{v,i}^2}} \; .
\end{equation}

Note that this asymptotic expression is independent of the choice of
$\mathbf{a}$.  Here, for large $M$, the benefit of having multiple
antennas at the FC hinges on the relative magnitude of
$M\sigma_{v,i}^2$ versus $d_i^{2\alpha}\sigma_n^2 $.  If
$M\sigma_{v,i}^2\ll d_i^{2\alpha}\sigma_n^2$, a reduction in variance
by a factor of $M$ is possible.  In this case, where the SNR at the FC
is low but the signals sent from the sensors are high quality, the
coherent gain from the combination of the relatively noise-free sensor
signals helps increase the SNR at the FC.  On the other hand, when
$M\sigma_{v,i}^2\gg d_i^{2\alpha}\sigma_n^2$, performance is
asymptotically independent of $M$.  Here, the coherent gain not only
applies to $\theta$ but also to the sensor noise, which is stronger in
this case.

\section{Impact of Imperfect Phase}\label{sec:five}

The previous sections have assumed that the FC can calculate the
vector $\mathbf{a}$ and feed the phase information back to the sensors
error free.  Whether the feedback channel is digital or analog, there
are about to be errors either in the received feedback at the sensors
or in how the phase shift is actually implemented.  Furthermore, the
wireless channel may change during the time required for calculation
and feedback of $\mathbf{a}$, so even if the phase shifts are
implemented perfectly at the sensors, they may no longer be valid for
the current channel.  In this section, we evaluate the impact of
errors in the sensor phase shifts on the estimation accuracy.

Define the phase shift for the $i$th sensor as $a_i=e^{j\alpha_i}$,
and assume that
\begin{equation}
\alpha_i=\alpha_i^{*}+\Delta_i \; , \nonumber
\end{equation}
where $\alpha_{i}^{*}$ is the optimal phase and $\Delta_i$ is a
Gaussian perturbation (in radians) with zero mean and variance $\sigma_p^2$.  Define
$\mathbf{E}=\mathbf{H}^{H}(\mathbf{H}\mathbf{V}\mathbf{H}^{H}+\sigma_{n}^2\mathbf{I})^{-\frac{1}{2}}$,
so that $Var(\hat{\theta}_{ML})$ can be expressed as
\begin{equation}\label{eq:var}
Var(\hat{\theta}_{ML})=\frac{1}{\|\mathbf{a}^{H}\mathbf{E}\|^2}=\frac{1}{\sum_{i=1}^M|\mathbf{a}^{H}\mathbf{e}_i|^2}\;,
\end{equation}
where $\mathbf{e}_i$ is the $i$th column of $\mathbf{E}$.  Let
$e_{i,j}e^{j\beta_j}$ be a polar coordinate representation of the
$j$th element of $\mathbf{e}_{i}$, so that
\begin{eqnarray}
|\mathbf{a}^{H}\mathbf{e}_i|^2&=&\left|\sum_{j=1}^{M}e_{i,j}e^{\alpha_j^{*}+\Delta_j+\beta_j}\right|^2\nonumber\\
&=&\sum_{j=1}^{M}e_{i,j}^2+\sum_{l=1}^{M}\sum_{\substack{m=1\\m\neq l}}^{M}e_{i,l}e_{i,m}\cos(\alpha_l^{*}+\Delta_l+\beta_l-\alpha_m^{*}-\Delta_m-\beta_m)\; . \label{eq:p1}
\end{eqnarray}
Define $\delta_{l,m}^{i}=\Delta_l-\Delta_m$ and
$\tau_{l,m}^{i}=\alpha_l^{*}+\beta_l-\alpha_m^{*}-\beta_m$.
If we assume $\sigma_p^2\ll 1$, (\ref{eq:p1}) may be approximated via
a 2nd order Taylor series as follows:
{\small\begin{eqnarray}\label{eq:err} |\mathbf{a}^{H}\mathbf{e}_i|^2\!\!&
\approx
&\!\!\sum_{j=1}^{N}e_{i,j}^2+\sum_{l=1}^{N}\sum_{\substack{m=1,\\m\neq
l}}^{N}e_{i,l}e_{i,m}\left(\cos(\tau_{l,m}^{i})-\sin(\tau_{l,m}^{i})\delta_{l,m}^{i}-\frac{\cos(\tau_{l,m}^{i})}{2}\left(\delta_{l,m}^i\right)^2\right)\nonumber\\
\!\!&=&\!\!\sum_{j=1}^{N}e_{i,j}^2\!+\!\sum_{l=1}^{N}\sum_{\substack{m=1,\\m\neq
l}}^{N}e_{i,l}e_{i,m}\cos(\tau_{l,m}^{i})\!-\!\sum_{l=1}^{N}\sum_{\substack{m=1,\\m\neq
l}}^{N}e_{i,l}e_{i,m}\left(\sin(\tau_{l,m}^{i})\delta_{l,m}^i\!+\!\frac{\cos(\tau_{l,m}^i)}{2}\left(\delta_{l,m}^{i}\right)^2\right).
\end{eqnarray}}
Substituting~(\ref{eq:err}) into~(\ref{eq:var}), we have
{\small\begin{eqnarray}
Var(\hat{\theta}_{ML})\!\approx\!\frac{1}{\sum_{i=1}^{M}\!\!\left(\!\sum_{j=1}^{N}e_{i,j}^2\!+\!\sum_{l=1}^{N}\sum_{\substack{m=1\\m\neq
l}}^{N}e_{i,l}e_{i,m}\cos(\tau_{l,m}^{i})\!-\!\sum_{l=1}^{N}\sum_{\substack{m=1\\m\neq1}}^{N}e_{i,l}e_{i,m}\left(\sin(\tau_{l,m}^i)\delta_{l,m}^i\!+\!\frac{\cos(\tau_{l,m}^i)}{2}\left(\delta_{l,m}^i\right)^2\right)\right)}.\nonumber
\end{eqnarray}}

In the previous equation, the effect of the phase error is
confined to the second double sum inside the outermost parentheses.
If we define $\hat{\theta}_{ML}^P$ to be the estimate obtained with
no phase errors, then
\begin{equation}
Var(\hat{\theta}_{ML}^P)=\frac{1}{\sum_{i=1}^{M}\left(\sum_{j=1}^{N}e_{i,j}^2
\!+\!\sum_{l=1}^{N}\sum_{\substack{m=1\\m\neq l}}^{N}e_{i,l}e_{i,m}\cos(\tau_{l,m}^i)\right)} \; ,
\end{equation}
which is deterministic and does not depend on the random phase
errors.  We can then obtain the following approximation
\begin{eqnarray}
Var(\hat{\theta}_{ML})\overset{(f)}{\approx}Var(\hat{\theta}_{ML}^P)\left(1+\frac{\sum_{i=1}^{M}\left(\sum_{l=1}^{N}\sum_{\substack{m=1\\m\neq l}}^{N}e_{i,l}e_{i,m}\left(\sin(\tau_{l,m}^{i})\delta_{l,m}^i+\frac{\cos(\tau_{l,m}^i)}{2}\left(\delta_{l,m}^i\right)^2\right)\right)}{\sum_{i=1}^{M}\left(\sum_{j=1}^{N}e_{i,j}^2\!+\!\sum_{l=1}^{N}\sum_{\substack{m=1\\m\neq l}}^{N}e_{i,l}e_{i,m}\cos(\tau_{l,m}^i)\right)}\right)\;,
\nonumber
\end{eqnarray}
where ($f$) is due to the first order Taylor approximation $(1-\frac{x}{y})^{-1}\approx 1+\frac{x}{y}$
for $x \ll y$.  We use the ratio of $Var(\hat{\theta}_{ML})$ to
$Var(\hat{\theta}_{ML}^P)$ to measure the effect of the phase error,
which yields
\begin{equation}
\frac{Var(\hat{\theta}_{ML})}{Var(\hat{\theta}_{ML}^P)}\approx \left(1+\frac{\sum_{i=1}^{M}\left(\sum_{l=1}^{N}\sum_{\substack{m=1\\m\neq l}}^{N}e_{i,l}e_{i,m}\left(\sin(\tau_{l,m}^{i})\delta_{l,m}^i+\frac{\cos(\tau_{l,m}^i)}{2}\left(\delta_{l,m}^i\right)^2\right)\right)}{\sum_{i=1}^{M}\left(\sum_{j=1}^{N}e_{i,j}^2\!+\!\sum_{l=1}^{N}\sum_{\substack{m=1\\m\neq l}}^{N}e_{i,l}e_{i,m}\cos(\tau_{l,m}^i)\right)}\right) \; . \nonumber
\end{equation}
Note that the only term in the above expression that is random
is the numerator on the right-hand side.

Taking the expectation of the ratio with respect to the phase perturbations $\Delta_i$,
we have
{\small\begin{eqnarray}
\mathbb{E}\left\{\frac{Var(\hat{\theta}_{ML})}{Var(\hat{\theta}_{ML}^P)}\right\}&=&\!\left(1+\frac{\sum_{i=1}^{M}\left(\sum_{l=1}^{N}\sum_{\substack{m=1\\m\neq l}}^{N}e_{i,l}e_{i,m}\left(\sin(\tau_{l,m}^{i})\mathbb{E}\left\{\delta_{l,m}^i\right\}+\frac{\cos(\tau_{l,m}^i)}{2}\mathbb{E}\left\{\left(\delta_{l,m}^i\right)^2\right\}\right)\right)}{\sum_{i=1}^{M}\left(\sum_{j=1}^{N}e_{i,j}^2\!+\!\sum_{l=1}^{N}\sum_{\substack{m=1\\m\neq l}}^{N}e_{i,l}e_{i,m}\cos(\tau_{l,m}^i)\right)}\right)\nonumber\\
&\overset{(h)}{=}&\left(1+\frac{\sum_{i=1}^{M}\sum_{l=1}^{N}\sum_{\substack{m=1\\m\neq l}}^{N}e_{i,l}e_{i,m}\cos(\tau_{l,m}^i)\sigma^2_p}{\sum_{i=1}^{M}\left(\sum_{j=1}^{N}e_{i,j}^2\!+\!\sum_{l=1}^{N}\sum_{\substack{m=1\\m\neq l}}^{N}e_{i,l}e_{i,m}\cos(\tau_{l,m}^i)\right)}\right)\; , \label{eq:p2}
\end{eqnarray}}where in ($h$) we exploit the fact that
$\mathbb{E}\left\{\delta_{l,m}^i\right\}=0$ and
$\mathbb{E}\left\{\left(\delta_{l,m}^i\right)^2\right\}=2\sigma^2_p$.
Since
\begin{equation}
\sum_{l=1}^{N}\sum_{\substack{m=1\\m\neq l}}^{N}e_{i,l}e_{i,m}
\cos(\tau_{l,m}^i)\le \left(N-1\right)\sum_{l=1}^{N}e_{i,l}^2 \; , \nonumber
\end{equation}
the ratio in~(\ref{eq:p2}) is approximately upper bounded by
\begin{equation}\label{eq:errbound}
\mathbb{E}\left\{\frac{Var(\hat{\theta}_{ML})}{Var(\hat{\theta}_{ML}^P)}\right\} \le
1+\left(1-\frac{1}{N}\right)\sigma^2_p \; .
\end{equation}
We see from~(\ref{eq:errbound}) that the impact of the phase errors
increases with $N$, but in all cases the degradation in the estimate
variance is approximately bounded above by a factor of $1+\sigma_p^2$.

\section{Sensor Selection}\label{sec:six}

As mentioned earlier, in situations where it is desired to use only a
subset of the sensors to estimate the parameter ({\em e.g.,} in order
to conserve power at the sensors), the FC needs a method to perform
the sensor selection.  Assuming only $K < N$ of the sensors are to be
selected for transmission to the FC, an optimal solution to the
problem would require solving the following maximization:
\begin{eqnarray}\label{eq:select}
\max_{\mathbf{a,x}} && \mathbf{x}^{T}\mathbf{D}^{H}\mathbf{H}^H\left(\mathbf{H}\mathbf{V}\mathbf{X}\mathbf{H}^H+\sigma_n^2\mathbf{I}_{M}\right)^{-1}\mathbf{H}\mathbf{D}\mathbf{x}\\
s. t. &&\sum_{i=1}^{N}x_{i}=K\nonumber\\
      &&x_{i}=\{0,1\}\nonumber\\
      &&|a_{i}|=1\;,\nonumber
\end{eqnarray}
where $\mathbf{D}=\textrm{diag}\left\{a_{1},\cdots,a_N\right\}$,
$\mathbf{x}=[x_1,\cdots, x_N]^T$ is the selection vector and
$\mathbf{X}=\textrm{diag}\{x_1,\cdots,x_N\}$.  Even if one chooses one
of the suboptimal approaches described in Section~III for estimating
$\mathbf{a}$, solving for $\mathbf{x}$ in~(\ref{eq:select}) requires
an exhaustive search over all possible $K$-sensor combinations and is
in general NP-hard.  Instead, in this section we derive conditions
under which much simpler selection strategies can be applied.  We
consider the following two cases: (1) low sensor noise relative to the
noise at the FC, $\sigma_{v,i}^2\ll\sigma_n^2$, and (2) relatively
high sensor noise $\sigma_{v,i}^2\gg\sigma_n^2$.  For (1), we derive a
LP solution as well as a simpler greedy algorithm, and
for (2) we show that the problem reduces to choosing the sensors with
the lowest measurement noise.

\subsection{Algorithms for High FC Noise}

Let $\mathbf{a}$ be the phase vector obtained using one of the
algorithms in Section~III assuming all $N$ sensors are active.  When
$\sigma_{v,i}^2\ll\sigma_{n}^2$, we ignore the term $\mathbf{HVXH}^H$
in~(\ref{eq:select}), and the problem simplifies to
\begin{eqnarray}
\max_{\mathbf{x}} && \mathbf{x}^{T}\mathbf{D}^{H}\mathbf{H}^H\mathbf{H}
\mathbf{D}\mathbf{x} \label{eq:select2} \\
s. t. &&\sum_{i=1}^{N}x_{i}=K\nonumber\\
      &&x_{i}=\{0,1\}\nonumber\;.
\end{eqnarray}
Define $\mathbf{F}=\mathbf{\mathbf{D}^{H}\mathbf{H}^H\mathbf{HD}}$
so that~(\ref{eq:select2}) can be rewritten as
\begin{eqnarray}
\max_{\mathbf{x}} && \mathbf{x}^{T}\textrm{Re}\{\mathbf{F}\}\mathbf{x} \label{eq:select3}\\
s. t. &&\sum_{i=1}^{N}x_{i}=K\nonumber\\
      &&x_{i}=\{0,1\}\nonumber\;.
\end{eqnarray}
Since $x_i^2=x_i$, (\ref{eq:select3}) is equivalent to
\begin{eqnarray}\label{eq:}
\max_{x_i} && \sum_{i=1}^{N}\mathbf{F}_{i,i}x_{i}+2\sum_{i=1}^{N-1}
\sum_{j=i+1}^{N}\textrm{Re}\{\mathbf{F}_{i,j}\}x_{i}x_{j}\label{eq:select4}\\
s. t. &&\sum_{i=1}^{N}x_{i}=K\nonumber\\
      &&x_{i}=\{0,1\}\nonumber\;,
\end{eqnarray}
where $\mathbf{F}_{i,j}$ denotes the $(i,j)\textrm{th}$ element of
matrix $\mathbf{F}$.  By linearizing the term $x_ix_j$
\cite{Billionnet:1997}, (\ref{eq:select4}) is equivalent to
\begin{subequations}\label{eq:intp}
\begin{align}
\max_{x_i,y_{ij}} & \sum_{i=1}^{N}\mathbf{F}_{i,i}x_{i}+2\sum_{i=1}^{N-1}
\sum_{j=i+1}^N\textrm{Re}\{\mathbf{F}_{i,j}\}y_{ij}\\
s. t. &\sum_{i=1}^{N}x_{i}=K\label{eq:linear0}\\
      &1-x_i-x_j+y_{ij}\ge0\label{eq:linear1}\\
      &x_{i}-y_{ij}\ge0\label{eq:linear2}\\
      &x_{j}-y_{ij}\ge0\label{eq:linear3}\\
      &y_{ij}\ge0\label{eq:linear4}\\
      &x_{i}=\{0,1\}\label{eq:linear5}\;,
\end{align}
\end{subequations}
where the constraints (\ref{eq:linear1})-(\ref{eq:linear5}) lead to
$y_{ij}=x_{i}x_{j}$.

Note that all of the constraints in~(\ref{eq:intp}) are linear,
except for~(\ref{eq:linear5}).  If we relax the constraint
in~(\ref{eq:linear5}), the condition $0\le x_i\le 1$ is implicitly
included in~(\ref{eq:linear7})-(\ref{eq:linear10}), and we are
left with a LP problem in standard form
\cite{Billionnet:1997}:
\begin{subequations}\label{eq:lp}
\begin{align}
\max_{x_i, y_{ij}} & \sum_{i=1}^{N}\mathbf{F}_{i,i}x_{i}+2\sum_{i=1}^{N-1}\sum_{j=i+1}^N\textrm{Re}\{\mathbf{F}_{i,j}\}y_{ij}\\
s. t. &\sum_{i=1}^{N}x_{i}=K\label{eq:linear6}\\
      &1-x_i-x_j+y_{ij}\ge0\label{eq:linear7}\\
      &x_{i}-y_{ij}\ge0\label{eq:linear8}\\
      &x_{j}-y_{ij}\ge0\label{eq:linear9}\\
      &y_{ij}\ge0\label{eq:linear10}\;.
\end{align}
\end{subequations}
To find the $x_i=\{0,1\}$ solution needed for sensor selection, one
can take the result of~(\ref{eq:lp}) and simply set the $K$ largest
elements to one and the rest to zero.  If desired, once the $K$ sensors
have been selected, the phase vector $\mathbf{a}$ for these $K$ sensors
can be recomputed based on a reduced dimension version of the algorithms
in Section~III.

The above LP problem has $\frac{N(N-1)}{2}+N$ variables and
$2N(N-1)+1$ constraints, and thus will require on the order of
$\left(\frac{N(N-1)}{2}+N\right)^2\left(2N(N-1)+1\right)$ arithmetic
operations \cite{Boyd:2004}.  A simpler greedy algorithm is presented
below that only requires $O(KN)$ operations, and that achieves
performance close to the LP approach.  The greedy algorithm is based
on the following observation:
\begin{eqnarray*}
\mathbf{x}^T\mathbf{D}^H\mathbf{H}^H\mathbf{HDx} & = &
\sum_{i=1}^K \sum_{j=1}^K \bar{a}_i a_j \mathbf{h}_i^H \mathbf{h}_j \\
& = & \sum_{i=1}^{K-1} \sum_{j=1}^{K-1} \bar{a}_i a_j \mathbf{h}_i^H \mathbf{h}_j
+ \|\mathbf{h}_K\|^2 + 2\mbox{\rm Re} \left\{
\sum_{j=1}^{K-1} \bar{a}_K a_j \mathbf{h}_K^H \mathbf{h}_j \right\} \; .
\end{eqnarray*}
The idea behind the greedy algorithm is to add sensors one at a
time based on those for which the last two terms in the above
sum are the largest.  The steps of the algorithm are detailed
below.

{\flushleft{\em Greedy Sensor Selection Algorithm}}

\begin{enumerate}
\item Select the first sensor as the one with the strongest channel:
$i = \arg \max_{k} \|\mathbf{h}_k\|^2$, and initialize the active
sensor set as $\mathcal{S}=\{i\}$\;.
\item While $|\mathcal{S}| \le K$, perform the following:
\begin{enumerate}
\item Solve
\begin{equation*}
i = \arg\max_{k\notin\mathcal{S}} \; \|\mathbf{h}_k\|^2 + 2\mbox{\rm Re} \left\{
\sum_{j\in\mathcal{S}} \bar{a}_k a_j \mathbf{h}_k^H \mathbf{h}_j \right\} \; .
\end{equation*}
\item Update $\mathcal{S}=\mathcal{S}\bigcup i$\;.
\end{enumerate}
\end{enumerate}
As with the LP algorithm, once the $K$ sensors are selected, an
updated solution for the associated $K$ elements of $\mathbf{a}$
can be obtained.

\subsection{Algorithm for High Sensor Noise}

When $\sigma_{v,i}^2\gg\sigma_n^2$ and assuming that $N > M$ (the case
of interest when sensor selection is necessary), the original criterion
can be simplified to
\begin{eqnarray*}
\mathbf{a}^H\mathbf{H}^H\left(\mathbf{HVH}^H\right)^{-1}\mathbf{Ha}
& = & \mathbf{a}^H\mathbf{V}^{-\frac{1}{2}}\mathbf{V}^{\frac{1}{2}}
\mathbf{H}^H\left(\mathbf{HVH}^H\right)^{-1}\mathbf{H}
\mathbf{V}^{\frac{1}{2}}\mathbf{V}^{-\frac{1}{2}}\mathbf{a} \\
& = & \mathbf{a}^H\mathbf{V}^{-\frac{1}{2}}
\mathbf{P}_{VH} \mathbf{V}^{-\frac{1}{2}}\mathbf{a} \; ,
\end{eqnarray*}
where $\mathbf{P}_{VH} = \mathbf{V}^{\frac{1}{2}}
\mathbf{H}^H\left(\mathbf{HVH}^H\right)^{-1}\mathbf{H}
\mathbf{V}^{\frac{1}{2}}$ is a rank $M$ projection matrix.  Ideally, to
maximize the criterion function, one should attempt to find a
vector of the form $\mathbf{V}^{-\frac{1}{2}}\mathbf{a}$ that lies in the
subspace defined by $\mathbf{P}_{VH}$.  Assuming the vector
$\mathbf{a}$ can approximately achieve this goal, the lower bound
on variance is approximately achieved and we have
\begin{equation}\label{eq:varlower}
\frac{1}{\mathbf{a}^H\mathbf{V}^{-\frac{1}{2}}
\mathbf{P}_{VH} \mathbf{V}^{-\frac{1}{2}}\mathbf{a}} \approx
\frac{1}{\mathbf{a}^H\mathbf{V}^{-1}\mathbf{a}} =\frac{1}{\sum_{i=1}^N \frac{1}{\sigma_{v,i}^2}} \; .
\end{equation}
With respect to the sensor selection problem, this suggests that
when $\sigma_{v,i}^2\gg\sigma_n^2$, the $K$ sensors with the smallest
values of $\sigma_{v,i}^2$ should be chosen.

\section{Simulation Results}\label{sec:seven}

Here we present the results of several simulation examples to
illustrate the performance of the proposed algorithms.  In all cases,
the path loss exponent $\alpha$ was set to $1$, and each result is
obtained by averaging over 300 channel realizations.  The sensors are
assumed to lie in a plane at random angles with respect to the FC,
uniformly distributed over $[0,2\pi)$.  The distances of the sensors
to the FC will be specified separately below.  To evaluate the
performance without feedback, $\mathbf{a}$ is set to a vector of all
ones.  In some of the simulations, we will compare the performance of
the proposed algorithms with that obtained by~(\ref{eq:amplitudeopt}),
where both the sensor gain and phase can be adjusted.  In these
simulations, we use the active-set method to
optimize~(\ref{eq:amplitudeopt}), and we use several different
initializations in order to have a better chance of obtaining the
global optimum.  When the ACMA algorithm is implemented, the
subspace dimension was set at $m=2$.

\begin{figure}
\centering
\includegraphics[height=3.7in, width=4.8in]{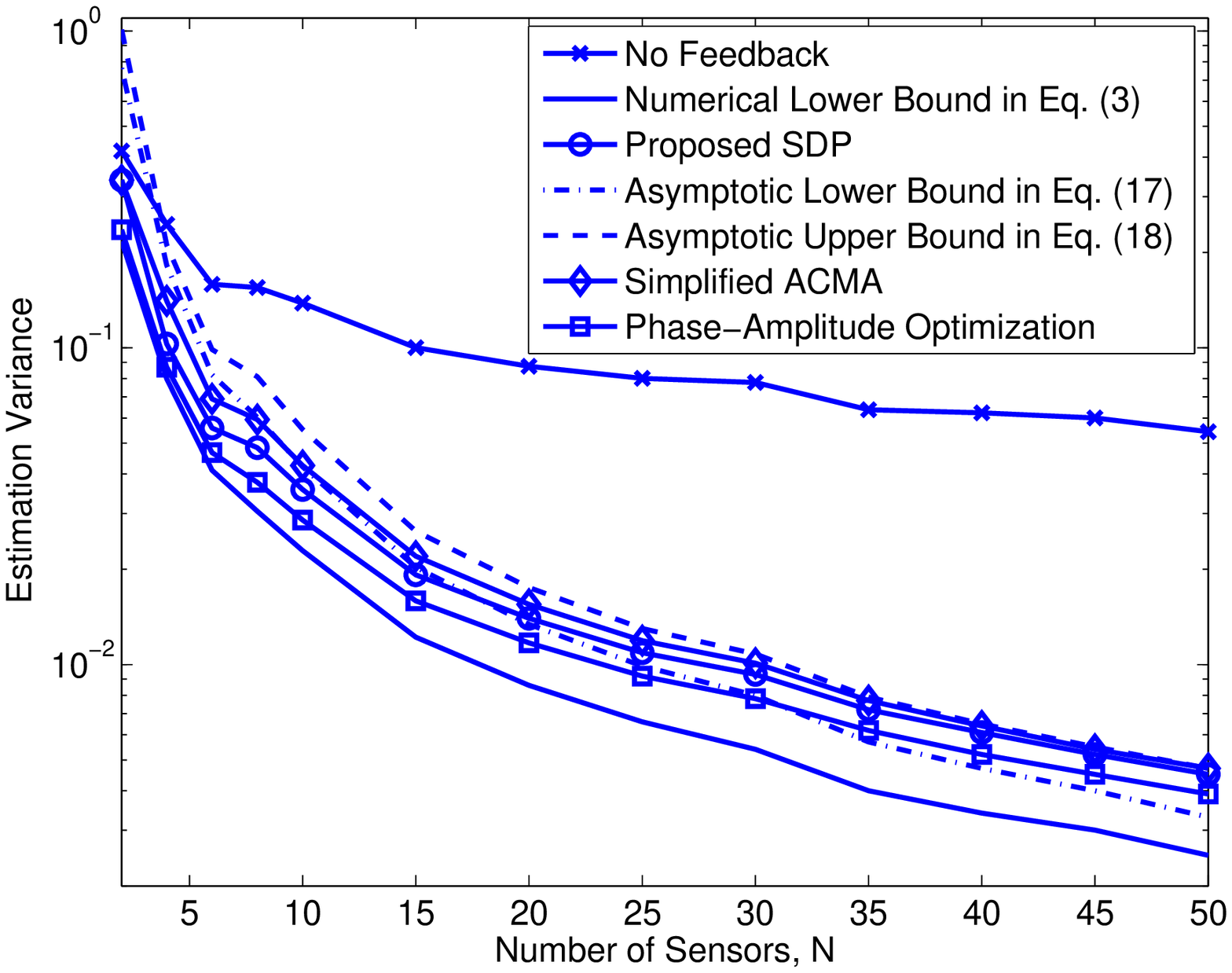}
\caption{Performance of the proposed algorithms with an increasing number of
sensors for a low measurement noise scenario ($\sigma_n^2=0.1$,
$\sigma_{v,i}^2$ uniformly distributed over $[0.01,0.1]$, $d_i$
uniformly distributed over $[3, 20]$ and $M=4$).}
\label{f1}
\end{figure}

\begin{figure}
\centering
\includegraphics[height=3.7in, width=4.8in]{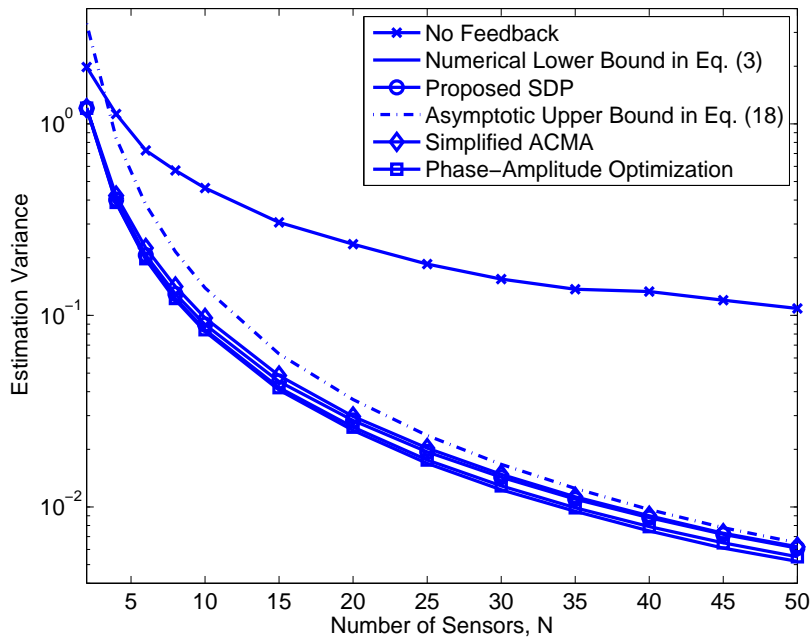}
\caption{Performance of the proposed algorithms with an increasing number of
sensors for a low measurement noise scenario ($\sigma_n^2=0.1$,
$\sigma_{v,i}^2$ uniformly distributed over $[0.01,0.1]$, $d_i=11.5$
and $M=4$).}
\label{f2}
\end{figure}

In the first two examples, we study the estimation performance for
$M=4$ FC antennas with increasing $N$ for a case where the sensor
measurement noise $\sigma_{v,i}^2$ is uniformly distributed over
$[0.01, 0.1]$ and the FC noise $\sigma_n^2$ is set to $0.1$.
Fig.~\ref{f1} shows the results assuming that the sensor distances
$d_i$ are uniformly distributed in the interval $[3,20]$, while in
Fig.~\ref{f2} $d_i=11.5$ for all sensors.  In both cases, even though
the lower bound of~(\ref{eq:lb}) is not achievable, we see that the
performance of the proposed SDP and ACMA methods is nonetheless
reasonably close to the bound, and not significantly worse than the
performance obtained by optimizing both the phase and gain.  As $N$
gets larger in Fig.~\ref{f1}, the estimation error for all of the
methods (except the no-feedback case) falls within the asymptotic
lower and upper bounds of~(\ref{eq:lb2app}) and~(\ref{eq:allone}).
When $N=50$, the ratio
$\mathrm{Var}\left\{\frac{1}{d_{i}^\alpha}\right\}/\mathbb{E}\left\{\frac{1}{d_{i}^{2\alpha}}\right\}$
is $0.304$ for Fig.~\ref{f1}, and the ratio between the lower and
upper bound is $0.702$, which is in excellent agreement with the value
of $1-0.304$ predicted by Eq.~(\ref{eq:ratio}).  Since the upper bound
in~(\ref{eq:allone}) corresponds to the case of $M=1$, one may suppose
that the gap in Fig.~\ref{f1} between the bounds of~(\ref{eq:lb2app})
and~(\ref{eq:allone}) indicates that the presence of multiple antennas
at the FC could provide a benefit for large $N$.  However, the
performance of SDP and ACMA is approaching the upper bound more
tightly, indicating that there is no benefit from having multiple
antennas in this case.  In Fig.~\ref{f2} where the $d_i$ are all
equal, the asymptotic bounds in~(\ref{eq:lb2app})
and~(\ref{eq:allone}) are identical, and asymptotically we expect no
benefit from multiple antennas at the FC.  We see again that for large
$N$ the performance of the SDP and ACMA methods is essentially at the
predicted bound.  When the $d_i$ are equal and
$\frac{\sigma_{v,i}^2}{d_i^{\alpha}}\ll\sigma_n^2$, the matrix
$\mathbf{H}^{H}(\mathbf{H}\mathbf{V}\mathbf{H}^{H}+\sigma_{n}^2\mathbf{I}_{M})^{-1}\mathbf{H}$
asymptotically approaches a scaled identity matrix, so in this case
the performance of the proposed phase-shift only algorithms even
approaches the lower bound of Eq.~(\ref{eq:lb}).

\begin{figure}
\centering
\includegraphics[height=3.7in, width=4.8in]{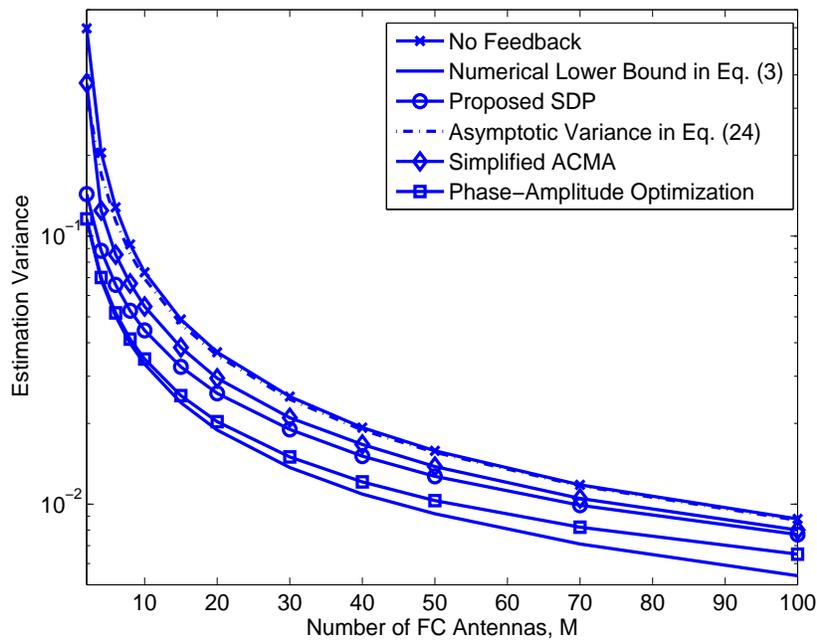}
\caption{Performance of the proposed algorithms with an increasing number of
antennas ($\sigma_n^2=0.1$, $\sigma_{v,i}^2$ uniformly distributed
over $[0.001,0.01]$, $d_i$ uniformly distributed over $[3, 20]$ and
$N=4$).}
\label{f3}
\end{figure}

Fig.~\ref{f3} illustrates the performance for $N=4$ with an increasing
number of FC antennas $M$ when $\sigma_{v,i}^2$ is uniformly
distributed over $[0.001, 0.01]$ and $\sigma_n^2=0.1$.  In this
example, for most of the sensors we have $M\sigma_{v,i}^2 \ll
d_i^{2\alpha}\sigma_n^2$, so in this case we see an improvement as the
number of FC antennas increases.  However, the benefit of optimizing
the transmit phase (and gain for that matter) is reduced as $M$
increases.

\begin{figure}
\centering
\includegraphics[height=3.7in, width=4.8in]{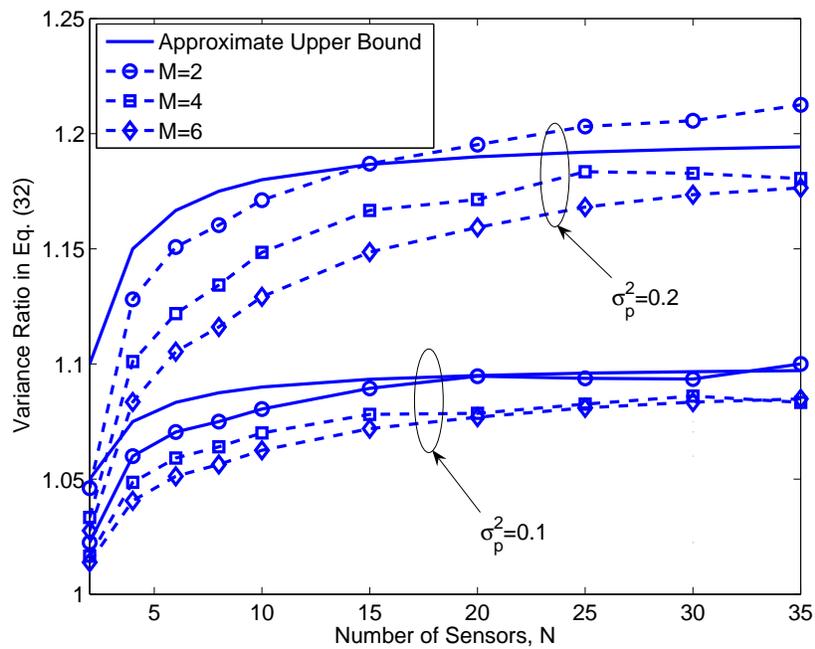}
\caption{Effect of phase errors on algorithm performance
($\sigma_n^2=0.1$, $\sigma_{v,i}^2$ uniformly distributed over
$[0.01,0.1]$ and $d_i$ uniformly distributed over $[3,
20]$).}
\label{f4}
\end{figure}

In Fig.~\ref{f4}, we investigate the effect of phase errors for two
cases, $\sigma_p^2=0.1$ and $\sigma_p^2=0.2$ assuming the same noise
parameter settings as in the first two examples.  For each channel
realization, results for 3000 different phase error realizations were
obtained and averaged to obtain the given plot.  The ratio of the
variance obtained by the SDP algorithm with and without phase errors
is plotted for $M=2,4,6$ for both values of $\sigma_p^2$, and the
approximate bound of~(\ref{eq:errbound}) is also shown.  The results
show that the performance degradation increases with $N$, and
that~(\ref{eq:errbound}) provides a reasonable indication of
performance for large $N$.  Fig.~\ref{f4} also shows that increasing
the number of FC antennas improves the robustness of the algorithm to
imprecise sensor phase.

\begin{figure}
\centering
\includegraphics[height=3.7in, width=4.8in]{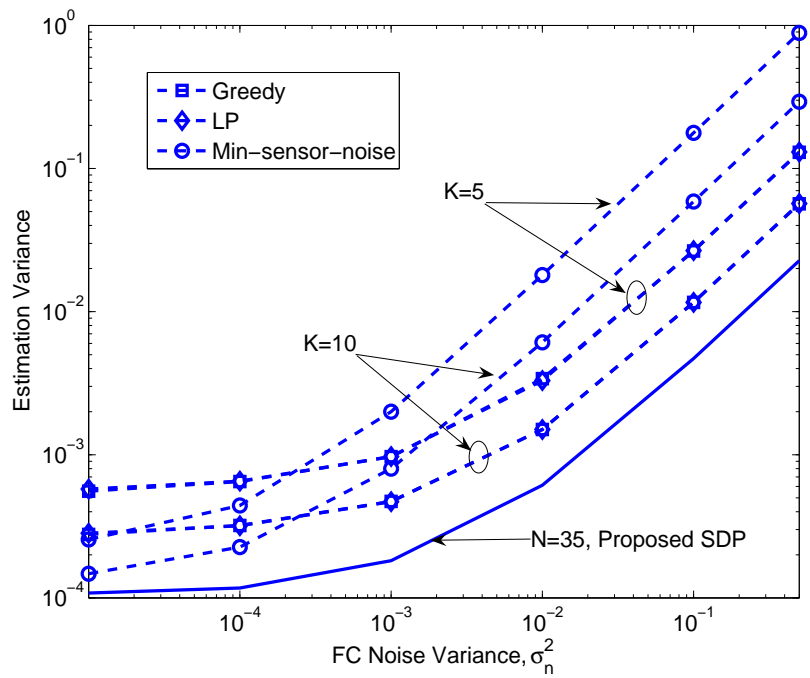}
\caption{Performance comparison between different sensor selection
algorithms ($N=35$, $M=4$, $\sigma_{v,i}^2$ uniformly distributed over
$[0.001,0.01]$ and $d_i$ uniformly distributed over $[3,
20]$).}\label{f5}
\end{figure}

In Fig.~\ref{f5}, we compare the performance of the three different
sensor selection algorithms discussed in the paper (LP, greedy and
min-sensor-noise) as a function of $\sigma_n^2$ assuming $M=4$
antennas, $N=35$ sensors and the sensor noise is uniformly distributed
over $[0.001, 0.01]$.  The sensor distances $d_i$ are uniformly
distributed in the interval $[3,20]$.  Three sets of curves are
plotted, one for $K=5$ selected sensors, one for $K=10$, and one
corresponding to when all the sensor nodes are used (the solid curve,
obtained using the SDP algorithm).  After the sensor selection, the
proposed SDP is used to re-optimize the selected sensor nodes' phase
parameters.  For small $\sigma_n^2$ such that $\sigma_{v,i}^2\gg
\sigma_n^2$, we see as predicted that the best performance is obtained
by simply selecting the $K$ sensors with the smaller measurement
noise.  On the other hand, again in agreement with our analysis, the
LP and greedy algorithms achieve the lowest estimation error for
larger values of $\sigma_n^2$.  Interestingly, the greedy algorithm
provides performance essentially identical to the LP approach at a
significantly reduced computational cost.

\section{Conclusions}\label{sec:eight}

In this paper, we investigated a distributed network of single antenna
sensors employing a phase-shift and forward strategy for sending their
noisy parameter observations to a multi-antenna FC.  We
presented two algorithms for finding the sensor phase shifts that
minimize the variance of the estimated parameter, one based on a
relaxed SDP and a closed-form heuristic algorithm based on the ACMA
approach.  We analyzed the asymptotic performance of the phase-shift
and forward scheme for both large numbers of sensors and FC antennas,
and we derived conditions under which increasing the number of FC
antennas will significantly benefit the estimation performance.  We
also analyzed the performance degradation that results when sensor
phase errors of variance $\sigma_p^2$ are present, and we showed that
for large $N$ the variance will approximately increase by a factor of
$1+\sigma_p^2$ provided that $\sigma_p^2 \ll 1$ square radian.  The
sensor selection problem was studied assuming either low or high
sensor noise with respect to the noise at the FC.  For low sensor
noise, two algorithms were proposed, one based on linear programming
with a relaxed integer constraint, and a computationally simpler
greedy approach.  For high sensor noise, we showed that choosing the
sensors with the smallest noise variances was approximately optimal.
Simulation studies of the proposed algorithms illustrate their
advantages and the validity of the asymptotic analyses.

\bibliographystyle{IEEEtran}
\bibliography{reference2}

\newpage

\end{document}